# A Re-Conceptualization of Online Misinformation Diffusion

Brett Bourbon
University of Dallas
bourbon@udallas.edu

Renita Murimi*
University of Dallas
rmurimi@udallas.edu

## Abstract

Online social networks facilitate the diffusion of misinformation. Some theorists construe the problem of misinformation as a problem of knowledge, hence of ignorance. This assumption leads to solutions in which misinformation (false belief) is resisted by good information (true belief). We argue that information is better understood as gossip. We believe that gossip spreads as part of an economy of social capital that has a specific discursive grammar that mimics ordinary human gossip. But there are some critical differences. These differences have immense and divisive social and political effects. If we shift our focus from the truth or falsity of information, and instead focus on the social dynamics of gossip we can more effectively respond to the challenges and dangers of online social networks.

Our argument has three parts. (1) We briefly critique epistemological and truth-centered accounts of misinformation. (2) We describe a basic discursive grammar of gossip as a social practice. (3) We, then, match the properties of online information with this discursive grammar of gossip. While gossip has a particular discursive form, its online modes involve a number of unique social features that will have immense and divisive social and political effects. Our goal is not to replace current accounts of information diffusion but to augment these accounts with a descriptive model of gossip. Information diffusion models should be understood as tools with which to explore the sociology of evolving online communities in conjunction with offline communities.

**Keywords:** misinformation, gossip, online social networks, information diffusion, fake news, viral spread

## 1. Introduction

Online social media is awash with misinformation, sometimes spreading at alarming rates. Some theorists construe the problem of misinformation as a problem of knowledge, hence of ignorance. This assumption leads to solutions in which misinformation (false belief) is resisted by good information (true belief). Swire-Thompson and Lazer (2020), for example, rely on an epistemological model of information, leading to proposals to counter misinformation by establishing epistemological legitimacy. While in some cases this epistemological focus is justified, evidence suggests that in the majority of cases it is not.

In two recent papers (Bourbon and Murimi, 2020, 2021), we have argued that information is often better understood as gossip, rather than knowledge. We believe that gossip spreads as part of an economy of social capital that has a specific discursive grammar (a kind of logical form) mimicking ordinary human gossip. If we understand the problem of information diffusion as not one about facts, data, or knowledge, but as the consequence of a peculiar form of gossip, we can more clearly determine how to respond to the challenges and dangers of online social networks.

Studying and developing better models of information diffusion remains of great interest to marketers and business, of course. But there is something of greater importance at stake. Social media enables the diffusion of information in various ways and through various channels. Nevertheless, what matters most remains people and their reactions and behavior with information. The sharing of information and the degree to which it diffuses through a network (or through multiple networks) forms and transforms the public spaces that have become a primary area of social interaction.

We will develop our argument in three parts. In Section 2, we offer a brief critique of epistemological and truth-centered accounts of misinformation. We provide a taxonomy of the varieties of misinformation, suggesting multiple ways of evaluating information separate from its veracity. In this first part, we reorient information away from misleading formal definitions and conceptualizations and towards an understanding founded in social practices and interpersonal dynamics. In Section 3, we will describe a basic discursive grammar of gossip as a social practice. This grammar will allow us to show that much of what we call online information and misinformation is conceptually equivalent to gossip. In Section 4, we match the properties of online information and the behavior of people using social media with the grammar of gossip established in Section 3. While gossip has a particular discursive form, its online modes involve a number of unique social features that will have immense and divisive social and political effects. Our goal is not to replace or criticize current accounts of information diffusion but to augment these accounts with a descriptive model of gossip. Information and misinformation diffusion should be understood relative to social dynamics and within the context of evolving communities, not relative to veracity.

## 2. The Varieties of Misinformation

The concept of information is both *equivocal* and *contested*. It is used to describe or label diverse and different things without recognizing this diversity and difference. As contested, the concept is often bound to various theories, such that arguing about what counts as information is tantamount to arguing about the merits of these theories. In such a situation what is required is not more theory, but minimal descriptions of kinds of information, that allows for a clarity of usage dictated by examples. In what follows we will begin to develop some minimal descriptions of not so much information and misinformation. Information and misinformation are multifarious and cannot be reduced to a single definition.

If we are to characterize misinformation we must distinguish it from its cousin disinformation. Disinformation could be construed as a kind of misinformation, but it is known to be wrong or false by those who spread it, and thus is a species of propaganda. Propaganda, which is a means of intentionally confusing or misleading people constitutes an irresponsible and dangerous practice, but that is not our target or concern in this paper. We are primarily interested in the spread of information as part of people's daily modes of interaction on social media. The difference between propaganda and political opinion, for example, will not always be clear, and at times that there is a difference will be denied by those of a different political persuasion.

The prevailing assumption is that information should be understood as true belief. When someone tells you that the light is on, and it is in fact on, then you have been informed of that fact. Your information is correct and thus it can legitimately be called by that name. The matching idea is often, therefore, that if the light is off, then you have been misinformed. That is true. But should we also conclude that this false information is not information at all, but instead a separate genus of statement called misinformation? Some claim that information must be true (veridical), if it is to count as information. The idea is that I am only informed if I am correctly informed. When I was incorrectly informed that the light was off, however, I was still informed of something. And when I realize I was misinformed, I was informed of something further (now I know, for example, that the light is *not* on). This is not simply a question of semantics. To claim that misinformation is information that is false implies something quite different from saying that there is something objectively defined as true information and something objectively defined as misinformation. In the second case, they belong to two genuses. In the first case, they are two species within the same genus.

Our ideas about the world, including our scientific ideas, change and alter over time. They fit with our other beliefs about nature, life, people, ourselves in complicated and often incoherent ways. Since our goal is to understand the flow and exchange of information in online environments within the context of our ordinary lives—that is we are studying living human beings through their interaction through technology—the relevant descriptions of information and misinformation will be grounded in examples. This also means that we want to offer descriptions that are minimally theoretical as possible. The theory of information put forward by Mingers et al (2018), for example, combines a host of theoretical commitments, including to a causal theory of truth, various arguments about the difference between a definition of information and a determination if something is or is not information, and so on. These theories come together in their claim that information must be objective and veridical to be information. For Mingers et al, signs are not interpreted; rather they "carry information whether or not they are observed" (87). We think this is incorrect, but instead of making philosophical arguments about the relationship between syntax and semantics, intentionality (Lower, 1997), externalism and internalism (Gates, 1996; Grice, 1957; Cummins, 1989; Soames, 2015; Brandom, 2010), and so on, we want to suggest that before we have elaborate theories we need examples. We do need a typology, but again what is needed is clarification about our everyday engagements with what we call information and misinformation.

We can side-step some of these theoretical issues, since they do not impinge on our particular problem. Our problem is the following: We want to understand everyday practices of information and misinformation exchange on the Internet. Thus, we do not need a logically universal description or definition of information. Since information is a vague term describing multiple things, it is not well-defined, and so such a logical universal description of information is theoretically hard to develop. (Of course, some researchers will want to remove the equivocation by nominating certain kinds of things as paradigmatically or essentially 'information.' Not only will this be question-begging if any one kind is taken as paradigmatic, it is irrelevant to our actual practices of information exchange, which involve many different kinds of information, including informing people of our emotional states and moods, and so forth.) What we need is a description of the ordinary and pervasive practices of information exchange and uptake by various online means.

A few important studies point to the inappropriateness of current models of information diffusion. An early study by Goel et al. (2012) shows clearly that information diffusion is seldom viral. While this does not directly support our contention that online information diffusion should be understood as a form of gossip, it does suggest that Internet behavior, while facilitated and constrained by network structures, is often more powerfully guided by the personal and social realities of which it is a part. The work of Tufekci & Wilson (2012) also shows how online behavior is both dynamic and embedded within offline human communities. Similarly, Vosoughi et al. (2018) discovered that false information spreads more rapidly and more deeply through a network than does true information. This again points to the critical role of social dynamics in information spread, where the social dynamics are best modeled as gossip.

We can characterize misinformation through a brief consideration of two recent research projects examining the structure of misinformation diffusion events. Goel et al (2016) returned to the question of how to determine if information cascades are viral a few years later. They again found that large online diffusion events did not conform to a decentralized viral model, but were rather better described as broadcast events, arising from spreaders who could broadcast the information or misinformation to their millions of followers:

*"It could have been, for example, that the very largest events are characterized by multi-generational branching structures–indeed that is the clear implication of the phrase "going viral." So, it is surprising that even the very largest events are, on average, dominated by broadcasts. It is also surprising that the correlation between size and structural virality is so low."*

We would like to suggest that two things are worth exploring relative to this observation. First, information is not self-replicating like a virus, and thus it is no surprise that large diffusion events lack structural vitality. Judgment and human decision are required, and those are not simply dependent on the information received, but also on the network of which they are a part. Second, the low virality of diffusion events of all kinds can be explained if information exchange is understood as part of the practice of gossiping. We gossip relative to our local concerns, which in general are defined by our friends and relevant acquaintances. Certainly, while social media extends the scope and number of gossip-friends, they still will be limited in number. And furthermore, one's friends do not simply get infected by the information we send them. If they did, then a viral event could easily be precipitated by those with a largest number of susceptible friends. Social dynamics, however, as we know, are not as simple.

The data from data-focused information diffusion studies do not enable us to discover the actual reasons, judgments, or psychological processes behind the spreading of misinformation. We have no evidence from the data that people believed the information that they spread. We can, however, define the range of possible belief attitudes towards the information and draw conclusions from these. Some spreaders may have known the information was false, and therefore spread it maliciously or as propaganda. Thus, they spread disinformation, and are a special case.

Those that remain, either thought the information was true or imagined that the truth did not matter. If we accept that they had no warrant to think the information true, then their belief in the veracity of the information and their spreading of it arose out of their motive for believing it true. For those who do not care about the truth, the same could be said. Those beliefs either motivated them to spread the information or not. If they did, then they were acting as a member of a belief community, sharing with believers or fighting against non-believers. Thus, in all cases, the information was spread as part of social purposes and dynamics.

When the putative misinformation is political, people of different political persuasions imagine that the solution to this kind of behavior would be to correct the beliefs of their opponents. This is unlikely to succeed and dangerous to attempt. This, of course, was the idea behind communist re-education camps, and the drive to eradicate false consciousness in those that did not realize the rectitude of the communist path (Cooke, 2019). The simple fact that follows from these considerations, however, is that OSNs have facilitated the creation of a hyper-gossiping rumor state. This also explains why there are so few viral events, and that often large diffusion events are media driven (associated with celebrities with high numbers of followers). You gossip primarily with your friends, especially those friends who have similar interests or commitments to the material shared.

Juul and Ugander (2021) reanalyze the data and conclusions of Vosoughi et al (2018) and Goel et al (2016) compensating for the size of diffusion events by using a method of subsampling. Their conclusions were mixed. They found that the fact-checked true- and false-news cascades are naturally similar:

*"Previously reported structural and temporal differences between true- and false-news cascades can be explained almost entirely by differences in cascade size, whereas the observed differences persist when comparing size-matched cascades of videos, images, news, and petitions".*

The authors conclude: "Our findings are consistent with the mechanisms underlying true- and false news diffusion being quite similar, differing primarily in the basic infectiousness of their spreading process". 'Infectiousness' in this context is defined by the authors as a high rate of diffusion. Since we think the mechanisms of information spread are not driven by epistemological goals, but rather by the social goals that motivate gossip, this similarity in 'mechanism' is what we would expect. The fact that false news spreads more rapidly requires explanation. Again, spreaders who knew it was false were either acting simply in bad faith or as propagandists. That is unfortunate. If we separate those people out, then we are left with everyday spreaders who either thought that the false news was true or they did not care or concern themselves with its truth or falsity. Even if these spreaders thought the information was true without adequate justification, that is not a reason to spread it more rapidly or eagerly, which is what Juul et al (2021) call 'infectiousness'. Something about the false news (more unusual or striking, maybe) motivated people to share. This does not mean they believed what they shared. We do not know. Whatever their motivations they were not centrally concerned with truth and falsity, but with social dynamics.

## 2.1. Taxonomy and Evaluative Rubric

Our proposed taxonomy for categorizing information does not delineate conceptually rigorous kinds. That would be inappropriate and hardly useful given the types of misinformation we generally have to deal with on the internet and in life. Our taxonomy is organized around the basic kinds of cognitive judgments we make when evaluating information: judgments of fact, interpretation, ideology, value, and scientific legitimacy. Each of these categories of judgement might involve the others. For example, facts must be be interpreted and that interpretation can be influence by ideology and so on. There are cases in which what matters most are ideological judgments as opposed to questions of fact. But at any time, this can shift.

With questions of fact, for example, we can make judgments about truth and falsity with greater confidence and agreement than we can about other more theoretical, interpretive, ideological, or value-laden claims. Even with questions of fact, however, we will have to interpret statements, and make judgments about context, salience, and so on. Nevertheless, we can delineate examples of information that turn not on fact, but on interpretive issues associated with human communication—intention, meaning or import, and significance. Issues of interpretation are also critical in reading and evaluating new reports, announcements, and even text communications with friends.

Interpretations are often shaped by ideology. Political polarization is so prevalent, however, that overt invocations of ideology organize much of what is called fake news and other forms of highly politicized misinformation. Questions of value can be involved in ideological commitments, but they extend beyond those. Value is a broad category—but the point is that how we evaluate information can be guided by our various values, not all of them ideological. Such values are much less susceptible to revision than are facts. Differences in values and attitudes towards life can cause people to profoundly misunderstand each other. It can make them evaluate information in radically different ways. And finally, we include questions of scientific understanding. As we have already commented, science is misunderstood when it is presented as producing truths. There are many different sciences, all of which have different standards of rigor, and so on. For the sake of simplicity, we have subdivided scientific information into accepted theories; applications of theories, and controversies. There is nothing definitive about such a categorization.

## 2.2. An Expanded List of Evaluative Judgements about Misinformation

We do not evaluate every instance of these various kinds of misinformation (or information) in the same way. We naturally want to evaluate assertions of fact as either true or false. But sometimes we can only say about any given statement that is probably true or probably false. For example, human behavior can be interpreted in multiple ways. Was the criminal motivated to commit his crime by poverty, hate, boredom, habit? There is no easy answer—and so we are left again making judgments about what is reasonable or unreasonable, likely or unlikely, and so on.

Another example. We might be tempted to call an ideology true or false, but relative to specific assertions that is often a confused gesture. Is it true or false that the United States is in decline? It depends on what you mean, and how you measure decline, and so on. Often it will simply be an expression of ideology (at least at first). So, one might say—that it is unfeasible to determine such a thing. Or we hedge our bets and admit the United States is incredibly rich, but our prestige is tarnished, and so on.

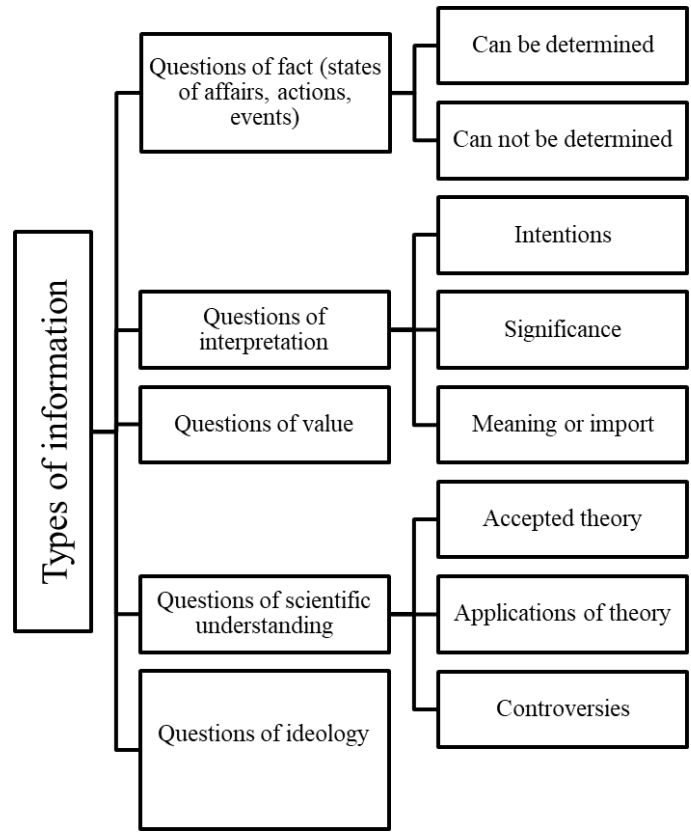

Figure 1. A taxonomy of misinformation

When a newsfeed sends out the information that an economist claims hot inflation will "wipe out 50% of the U.S. population"— some will shout fake news, others will say, true. But this is a predication, about which we might say it is probable or improbable. Similarly, pictures are posted on instagram, governments release statements on their websites, and celebrities express various political sympathies through their twitter accounts. All of these in point of fact might have no factual errors, but they can still be highly tendentious and misleading. A brutal crackdown, for example, is called 'a return to law and order.' A photograph, even if it has not been altered in any way, suggests something misleading. We must interpret, appealing to reasonableness, possibility, probability, feasibility and so on.

There is nothing surprising in this. It is a reminder. The issue with information is not always about its truth or falsity. Judgements about possibility, probability, and reasonableness can be as difficult to adjudicate as judgments of truth. People often confuse what is possible with what is probable, leading to all sort of mistaken claims about people, situations, and events. Nevertheless, we think a carefulness about kinds of misinformation should be matched with a conceptual or discursive care with our terms of judgment. This is especially the case if much of what we call information or misinformation is better described as gossip. We shall argue that this is so in the remainder of the paper.

# 3. A Discursive Grammar of Traditional Offline Gossip

Our goal in this paper is to show that what we call information and misinformation diffusion is akin to gossip. Gossip is a discursive practice, and is not simply a kind of statement with a particular kind of content. It is a social practice. Hence, in order to understand what gossip is we need a description of that practice. As a social practice, gossip consists of an ill-defined range of discursive modes, from first person observations and explanations to vague hearsay. Gossip, therefore, we might say, has both a social and a discursive grammar. It is something we say, but to understand it is to understand the practice of doing it. (This is again why it lacks anything like a logical form.)

The development of such a discursive grammar, however, is not an inductive research project. The goal is to derive a description of the essential normative characteristics of gossip as practiced within the communities of which we are a part. Those characteristics might be modified, by the further extension of the grammar to cover cases not at first envisioned. The evidence from which the grammar of gossip can be built is just what any competent speaker and social actor within a community would recognize as gossip. There will be liminal cases, of course. Relative to such cases, the ordinary discussions that language speakers might have about what counts as the normative sense of gossip would be what the theorist would also pursue. This makes the description of gossip contingent on the normative practices within any community. But those practices are in fact what we are trying to understand, not some theoretical, putatively necessary description, the relevance of which could only be determined by examining actual practices and modifying that definition relative to those practices. And that is tantamount to replacing a logical description with a discursive description of practice. Unlike the early work of Allport and Postman (1947), our goal is not to understand the psychology behind gossip, but to describe the discursive logic (its grammar) of the practice so that we can compare that logic to the practices of online information exchange.

## 3.1. A Structure of Gossip

The practice of gossip can be characterize using four characteristics, which we will call patterns:

I. Irresponsible Speech
II. Unwarranted Assertion
III. Exclusion of target of gossip
IV. Masked Implication and Import

These four patterns describe the differences between gossip and ordinary face-to-face discourse. We should all be able to recognize these patterns.

### Pattern I: Irresponsible Speech

In gossiping, we exchange impressions. We share things we see, or think we see. We do not determine what is actually happening, we simply say things like—I saw him park his car near her apartment, suggesting in saying that he is, therefore, visiting her. What we say may or may not be true. Similarly, we gossip by reporting what we have heard or what we have heard as hearsay. We say: 'Manager x said this to y.' Maybe manager x did say this or maybe she did not. Regardless, we pass it on.

While gossip can have specific origins, e.g., in things seen by a particular person, no one is understood as responsible for what is said. Gossip is not speech for which anyone stands as guarantor. One says, e.g., "I am just telling you what I saw" or "this is what people are saying." Anyone can speak gossip, because no one need be responsible for what is said or meant. In gossiping, it is as if we are exchanging quotations or our impressions of something, and thus we are simply saying 'This is what it seems'. When we gossip we act, or pretend to act, not as speakers or agents, but as vehicles. This does not mean that we are actually simply vehicles of information exchange, but we have not confirmed what we are sharing. We are not speaking as authorities separate from the quasi authority that we have seen or heard something (even if second, third, or fourth-hand).

Since we often act as if no one is responsible for what is said (and we are often vague about what might be meant by what is said), we do not often question the veracity of gossip. We do not determine its warrant (at least not rigorously or in good faith), especially if it fits what we want to believe or confirms our assumptions. All of this has the effect of lowering our resistance to sharing what we have heard, thus, facilitating its spread. Consequently, if no one is really responsible, then we can use what we hear for our own purposes; we react to it irresponsibly. Our self-interest and our self-doubts become sovereign in our interpretations and responses.

**Pattern II: Unwanted Assertion**

We often gossip about things about which we know very little., or we gossip about something for which we have no adequate reasons for belief. Consequently, gossip is talk for which we lack adequate warrant or justification; hence it is a form of hearsay.

Gossip is not part of a practice of epistemological evaluation. Rather, it is part of a social economy in which beliefs and impressions are exchanged. It constitutes an economy because these exchanges are acts of social significance. Sharing gossip indicates or encourages an intimacy, a sense of trust, and thus can encourage a kind of alliance, even if momentary. Only when the gossiping becomes reciprocal and frequent is the social relationship (alliance) fully established. This is similar to grooming practices in chimpanzee groups (See R. Dunbar for an evolutionary theory about this). Consequently, people arbitrate and evaluate gossip relative their beliefs and anxieties about things. As a consequence, it encourages facile belief, partly because of its lack of warrant. Gossip confirms prejudices more readily than it counters them.

Because gossip tends to be evaluated relative to beliefs and anxieties rather than to states of affairs, it is difficult to counter by evidence or counter-statement. (This does not mean that it is not about states of affairs, only that people's beliefs and anxieties about something matter more in how they respond, take up, and repeat gossip). Because of this, one fights gossip with more gossip, not with facts or rational inquiry. By filling the channels of social gossip with new gossip and counter-gossip one creates noise and alternate descriptions. Gossip displaces gossip in the way that Hume argues affections displace affections (and for a similar reason). This is an important fact to remember when thinking about the role and forms of gossip online.

We now summarize the discursive grammar of gossip at this point. Because no one is responsible for what is said and meant, gossip exists in *the exchange and sharing.* Gossip can sometimes be true, but its truth is nothing to believe, since you have no warrant to believe it (yet). In the best case, gossip is unjustified true belief. Even when its source is known, gossip functions as such because no one is responsible for its particular claims. It circulates through a self-reinforcing mode of exchange: no one speaks in their own voice, they simply pass the information on. Gossip understood relative to these two disjunctions becomes more of an event that happens to people than an action that people do; and so, people take it as something in the world, a kind of fact to react to as an opinion to dispute.

### Pattern III: Exclusion of Target of Gossip

Gossip is often defined as a conversation in which the target of that conversation is absent and cannot, therefore, take part in it. This definition is misleading. I can tell a friend of mine what people are saying about her. In so doing, I share with her a bit of gossip about her. She is present. Her absence is not required for me to share this gossip. What is required, however, is her silence within the gossip conversation. She can protest, but that is outside the gossiping. She would be making a comment about the gossip. Roland Barthes, in *Lover's Discourse*, characterizes the situation by arguing that the target of gossip is reduced to third person status, denied her first person or second person powers of speech or self-description. She has, in effect, no voice in the conversation. She has become a spectator, even if a protesting one.

In gossiping, people often share secrets, and in so doing they share private concerns in a public forum. The sharing of privacy, usually of someone else's privacy, encourages a sense of intimacy with those with whom we share the gossip at the expense of the privacy of the target. The consequence of the exclusion and silencing of target removes one source of resistance to what is said about that person. The target of the gossip is often treated as if they were mere tokens in some larger game or as it they were simply objects of our voyeuristic fascination.

The exclusion (silence or silencing) of the target reinforces the community of gossipers at the expense of the silenced or silent target. This exclusion reinforces the inclusion of those who gossip into a kind of discursive community. This community may be very short-lived. Gossip, therefore, facilitates the formation a community of intimacy, focused on a common friend or enemy.

### Pattern IV: Masked Implication and Import

When we gossip, we are not taking responsibility for what we say. As seen in the pattern of irresponsible speech, gossip can allow us to mask our purposes when we share it. We seemingly highlight what we see or the literal sense of what is said, leaving the implications for others to interpret. Consequently, we can use gossip to deceive by implication, protected behind a mask of innocence and obviousness. If I say, I saw Robert park his car near the house of 'You-know-who', I am not just reporting an observation. I am making a suggesting: implying a secret assignation. In this case, I am pretending simply to share a bit of information, telling you just what I saw. But what I mean is not quite what I say. Gossiping encourages us (and allow us) to understand statements relative to the inferences we draw from them. In other words, gossip as a social practice is often governed by a concern for secondary effects. Gossip prompts the imagination; it insinuates, implies, triggers associations, casts aspersions.

## 3.2. A General Picture of Gossip

We can collect these descriptions of gossip into two fundamental characterizations:
Gossip is more expressive of the community of gossipers than it is about the target of the gossip. What is said in gossip, because of the irresponsibility in which it is exchanged and its lack of warrant, and because of the way gossip is dispersed and held within the community of gossip exchange, floats free from responsible counter-statement.

While gossip is often about states of affairs, it also invokes and provokes desires, vanity, distress, and anxieties. Gossip has no foundation or warrant that can be readily examined; it is by definition irresponsible speech, and thus bounded only by its transmission and the dispositions of a community to accept it. Small face-to-face communities, especially within cultures oriented toward honor and reputation, police themselves through gossip; a kind of social mode of governance.i The volatility of gossip is checked by the face-to-face nature of such communities, and limited by the stricter proprieties organizing life within clans and families. When breakdowns do happen in such communities they tend to be violent and extreme. See, for example, Fanshen, William Hinton's remarkable historical account of the breakdown of the traditional order in a Chinese village during the Cultural Revolution. Propriety, a normative pattern of behavioral constraint, infused with moral overtones, dampens the disruptive powers of gossip, which is often motivated by resentments and directed toward particular targets. When propriety breaks down through social change or ideological fervor, as in the Cultural Revolution, gossip becomes an unchecked form of social revenge and power. Similar cultural conditions have been created by OSNs, leading to similar breakdowns in propriety facilitated by gossip.

One check on gossip in face-to-face communities as we have mentioned is social propriety, an accepted set of norms and standards of behavior governing everyday interactions among people within a community. Proprieties, even if misguided, express forms of respect. They constrain oneself as much as they can constrain others. Their manipulation can be devastating for a community because of this. Gossip expresses forms of authority in the way that secret knowledge or propaganda can. Indeed, gossip is a form of propaganda, while proprieties are forms of custom. Proprieties can be abused and they tend to conserve inequalities. Gossip, as Tocqueville rightly understood, forms the context for egalitarian societies, which are stabilized socially and relative to individual anxiety by a tendency for individuals to fit themselves within mass beliefs, within forms of social gossip as opposed to structures of propriety.

Gossip and propriety constitute two complementary aspects of an interlocking social economy. However, gossip, as practiced in small face-to-face communities, differs from the type experienced in online settings. Modern media, egalitarian politics, consumer capitalism, and the internet all facilitate gossip and rumor in a public mode, but without the limits of a face-to-face society. In addition, one dangerous aspect of OSNs is that they further dissolve the proprieties that are already diminished in modern society. Trolling and other aggressive forms of behavior have become ubiquitous in the anonymous and mediated forms of expression allowed on the internet. This decay of propriety and unleashing of solipsistic aggression feed into the naturally irresponsible forces of gossip. The checks remain those Tocqueville recognized: the natural stabilization and constraint provided by public (majority) opinion and sentiment. The effect of these checks are balkanized on the internet, however, since people tend to interact within mutually self-confirming, often exclusive communities.

# 4. Social Media as a Gossip Economy

## 4.1. A General Picture of Gossip

As one of the most important transactional spaces, social media becomes a mode not simply of communication, but a means and a context for the formation of communities and the exchange of gossip. Since communication is highly mediated by the Internet and its social platforms, information easily falls into the logical form of gossip, where no one is fully responsible for its content, and thus warrant is not relevant. Information becomes more rumor than fact. The easy proliferation of information and the speed and superficiality of our online interactions encourages often highly prejudicial and unthoughtful transfers of putative information; again, more than fact. In the next section, we will develop this general description into a more detailed correlation between the discursive grammar of gossip with our OSN interactions.

## 4.2. Detailed Analysis

In this section, we identify specific similarities between the discursive grammar of gossip we have developed above and the patterns of behavior characteristic of information exchange within OSNs. We have organized our analysis into eight sections, within which we show how OSNs facilitate information diffusion as part of a gossip economy.

[1] The tendency of people to befriend others who seem to share similar beliefs and commitments and the design of social media algorithmic push-strategies creates a situation in which the targets of negative statements are absent and silent relative to those making or sharing these statements (regardless of whether these statements are true or false). This is a prime characteristic of gossip, which we identified as the third pattern in the grammar of gossip (the exclusion of the target of gossip). Such a situation is not only self-confirming, but it reinforces the group of identity of those who gossip against those who are the absent targets of this gossip. Bessi et al (2015) in their study of Italian Facebook, for example, found that communities supporting various conspiracy theories formed easily, leading to the formation of social "echo chambers."

[2] Many social media theorists construe knowledge as personal belief, and classifying all information as data. This is misleading. Much that passes for informational data (hence as evidence) remains highly tendentious and untested. Similarly, personal belief hardly counts as knowledge, since knowledge must be not only believed, but justified and true. (It is true that the traditional analytic definition of knowledge as justified true belief does not cover all cases, as Gettier (1963) was first to demonstrate with his famous thought-experiment. It remains true, however, that if I cannot justify what i think I know, what I think I know is prejudice or accepted opinion.)

Online interactions encourage factitious justification, too often tied to the perceived status and authority of the various people involved. The proliferation and display of putative credentials, certifications, and other gestures towards authority facilitate the acceptance of misinformation as true. Confirmation bias and a reticence to question one's own beliefs encourages an uncritical attitude towards acceptable information and an unbalanced critical attitude to unacceptable information. Consequently, our highly mediated interactions through OSNs often take the form of personal advertisements and displays of status. What is advertised and sold through OSNs are the means of confirming our beliefs. Information accepted without adequate justification manifests the second pattern characterizing gossip (unwarranted assertion).

[3] Gossip encourages belief without adequate warrant. People with little understanding of science or statistics, for example, accept journalistic statements about scientific studies, without realizing how provisional such studies often are. Scientists are not immune from this. They, too, often draw erroneous conclusions and make numerous statistical errors, as well as being subject to confirmation bias and the negative effects consensus (about statistics, see Reinhart, 2015; Spirer et al, 1998; about consensus errors in science, see O'Conner and Weatherall).

In non-scientific online communities, the exchange of information encourages superficial kinds of knowledge (often simply pseudo-knowledge). This information lacks context, and people often have little sense of the questions and controversies that inform any particular intellectual discipline. In addition, information is read (or 'consumed') amidst people's other activities. The prime activity on social media, besides passive consumption of various stimuli, is browsing, which is a kind of stimulus-response activity, the primary effect of which is entertainment, which tends again towards easy confirmation of previous beliefs or basic ideological commitments (Vitak et al, 2011). This matches the second pattern characterizing the grammar of gossip (unwarranted assertion).

[4] Online transactional spaces like Facebook and Yahoo Homepage create the illusion of a neutral environment of information. This has the effect of diminishing the sense that anyone in particular is responsible for the information offered. This is one of the central characteristics of gossip as hearsay: no one is responsible for its content. This conforms to our first pattern of irresponsible speech. We see a similar situation in the exchange of online information, encouraged by the 'Like button' (and negatively by trolling). People naturally want other people to like their posts. Liking a post might express agreement, but its primary effect is to mark an exchange or investment—an investment of social capital through a transaction of praise (or in other contexts of blame). People collect positive (or negative) transactions. Such simplified interpersonal transactions mimic gossip's social function of establishing and maintaining allies and communities, creating an immediate and vast means of establishing social value (monetized in what are called 'influencers').

[5] Social dynamics drives and sustains the exchanges of information online. We have already noted three aspects of information diffusion that suggest (they do not prove) the primacy of the social dimension of this diffusion. (1) Information is not self-replicating. It is spread by choice and judgment, relative to social value. This is why most diffusion events, as Goel et al (20212; 2016) discovered, are neither viral nor large. (2) Information diffusion is primarily rumor, and rumor must be sustained by a community. (3) Our taxonomy of misinformation also highlights the social usages of information, and its contingent roles within our online and offline social reality.

[6] Truth and falsity are not the primary drivers for information spread, but rather personal and group investment in certain kinds of information are the primary drivers of this phenomenon. We have discussed in Part 1 how research into the spread of misinformation by Vosoughi et al. (2018), Goel (2012; 2016), and Juul et al (2021), in effect, support the critical role of social dynamics in information diffusion. The value of information is often not epistemological but social.

The motives for sharing misinformation on social media have been investigated by Chen et al (2015). They have also found that social factors play a primary role in the spread of information. For example, provocative information that prompted online conversations spread more easily than less catchy information. Similarly, social and interpersonal motives play a critical role in misinformation diffusion. Laato et al (2020) studied these social factors in the spread of COVID-19 misinformation, including the belief that 5G network towers facilitated the contagion. They found evidence of what others have called cyberchondria: the "unfounded escalation of concerns about common symptomology based on review of online content" (Starcevic and Berle, 2015). This fits with the general description of gossip defined above: gossip is more expressive *of* the community of gossipers than it is *about* the target of the gossip.

[7] The primary effect, if not purpose, of such gossip is the building and confirmation of a community of gossip-friends, often in relation to social enemies or outliers. What is shared or produced by such gossip-friends are rumors, and not information modeled on knowledge. It does not count as justified true belief. Justification is lacking and its truth is often irrelevant. Again, this conforms to general characteristic of gossip as expressive *of* the community of gossipers.

i. What we are calling gossip-friendship characterizes the communities of diffusion in OSNs. Gossip-friends share a propensity to exchange gossip online about certain topics. Such online friendships will in general be contingent and limited (although they need not be). Online relationships, while they follow the basic degrees of intimacy we see in offline relationships, are far more replaceable and labile. (Arnaboldi et al (2015). Twitter networks, in particular, involve a high turnover of connection when compared to offline networks. Arnaboldi et al (2015) explain this as both a consequence of fewer family connections in online networks and social opportunism, encouraging a "social butterfly" attitude towards social friendships.

Gossip-friendship creates the possibility for a new Word-of-Mouth (WOM) model of information diffusion. Online gossip diffusion often follows a pattern like word of mouth diffusion, but with an important difference. WOM "communication takes place within a social relationship that may be categorized according to the closeness of the relationship between information seeker and the source, represented by the construct tie strength" (Money et al. 1998; Duhan et al. 1997; Bristor, 1990). Online gossip follows similar patterns, but it is less dependent on "the relationship between information seeker and the source." Or rather that relationship need not be of any particular intimacy, either emotionally or as reflected in frequent communications. What is required is trust. Trust is personal—and thus often given for inadequate reasons. It can be motivated by an attraction to fame and personality or by ideology. Celebrities are prime figures in this, since they are not part of a network of reciprocal communication. Their influence is asymmetric, but the disproportionate level of trust afforded to them allows their statements to spread in similar ways to other word of mouth diffusions. (There are variations in the stability of online relationships depending on the users 'commitment and interest in using a particular platform. There are, for example, aficionados of Twitter, as well as regular and casual users).

In WOM networks, trust and intimacy are correlated. Information in such networks flows through such correlated ties. Research suggests that tie strength affects information flows. Brown & Reingen (1987) found that "Individuals in a strong tie relationship tend to interact more frequently and exchange more information, compared to those in a weak tie relationship." This is a study that predates the internet. Our hypothesis is that some OSN will replace intimacy with ties that imply trust (and that are not reciprocal, as with influencers), but that the degree of trust of these OSN will track WOM information diffusion, despite the difference in intimacy. Brown et al (2007) offer some evidence for exactly this effect. OSN in effect become parodies of offline social interaction, highlighting the most volatile aspects of our social relationships.

Gossip friends exist within the specific possibilities enabled by OSNs. One can see this in the interactions facilitated by TikTok and Instagram. Both of these platforms highlight the power and effects of celebrities and so-called influencers. Their power of influence, however, has been absorbed into the gossip structure of these platforms. These platforms simultaneously reward fame and flatten the difference between these influencers and everyone else. (The network is flattened because hierarchical differences between nodes are diminished; one can have a greater degree of connection and closeness online with influencers than one can have offline, e.g.). On both platforms these influencers respond to their fans. People compete for attention. On TikTok in particular the difference between a regular user and someone influential is simply an effect of popularity and marketing effects. Celebrities in the traditional sense of famous people who seem somehow separate and well-known, however, still retain their power. They have huge followings. But the putative intimacy of the internet, so much greater than that produced in earlier forms of media propaganda, allows such figures to function not as intimate friends of any kind, but as gossip friends, fitting within a network of influencers and various other kinds of online friendships. What this means is that celebrities and influencers, who would not be part of any real friendship network, expressive of real mutual connection or as manifest in a high rate of communication, function online as gossip-friends. (This is not a democratization of the social order, but a false intimacy that gives celebrities and influencers greater power of effect).

One new quality of these kinds of gossip-friendships is that they need not be reciprocal and can be asymmetrical. On TikTok influencers (or their handlers) can respond to comments. Anyone is a potential celebrity. But with more global celebrities, their influence is all towards consumers, but the relationship is much less emotionally distant than was likely in pre-internet media.

[8] Because the Internet encourages a breakdown of proprieties and manners, it skews its information economy towards gossip, with little of the face-to-face checks that traditionally mitigate the negative effects of gossip (unwarranted or unjustified hearsay). In Zollo et al (2015), the authors studied the emotional responses evoked by misinformation. They found that with an increase in the length of online discussions, characterized by an increase in the number of comments, the sentiments of posts became increasingly negative, as we have shown with the fourth pattern characterizing gossip (masked implication and import).

These eight points suggest that our OSN behavior should often be described as a form of gossiping. Although, the Internet can facilitate learning in specific and constrained domains, what we call information is best understood relative to the communities it helps form and maintain. Online exchanges might ultimately lead to our acceptance of information as knowledge, but often the factual questions are absorbed in social dynamics. Information must be understood not simply relative to its content, but relative to its role in a social economy. News feeds, like on Facebook, for example, since they are themselves filters reinforcing the interests and beliefs of particular user, are always in danger of becoming rumor-mongering channels. The exchange of gossip has social significance, a mix of competition and cooperation for social advantage. The truth or falsity of what is exchanged matters primarily relative to social advantage (or disadvantage). The truth value of the gossip will often be irrelevant to this social value. The upshot of this is that when we model information (or misinformation) diffusion we are not modeling (or trying to explain) something called information. We are attempting to model the actions and behavior of people through the effects and traces of that behavior on online social networks.

## 5. Conclusions

The internet, as the host to online social networks, despite its benefits, remains a dangerously transformative technology. It is an interconnected system of computer networks, a communication device, a multipurpose tool, a global infrastructure. We might understand it by analogy, therefore, with the invention and implementation of telephone networks or like the development of affordable automobiles and of the road systems to accommodate them. The internet's transformative effects, however, extend beyond these analogies. Its ascendency has been far faster than these earlier technological incursions. The telephone and the automobile, in combination with other technologies and social and economic change, opened up new ways of living and communicating. The internet, like the computer, is a multipurpose tool, and so we find uses for it in more and more aspects of our lives. The telephone and the automobile extend the natural powers of human beings. Human use and involvement with the internet, however, reshape the fundamental ways in which we organize ourselves into communities and understand ourselves as individuals. In this sense, we might analogize its growth and emergent effects as similar with the effects that follow from the development of cities that replaced the social organizations of Neolithic villages. We are currently living amidst a fundamental cultural change the consequences of which we have yet to fathom. But even if we live in this new Online Social Network city, our behavior and cultural understanding remains village bound. Online social networks, for example, facilitate gossip and crowd behavior endemic in earlier forms of village life (think of the novels of Jane Austen), but now universalized and made anonymous, without the face-to-face constraints organizing earlier forms of community. People's attractions to gossip are primal, but the means of expressing these attractions now functions through technology that augments the power and scope of our gossip behavior. This is a volatile situation.

Online social networks mirror, augment, and feed into the human propensity to gossip. We have argued that gossip has two fundamental characteristics:

a. Gossip is more expressive *of* the community of gossipers than it is *about* the target of the gossip.
b. What is said in gossip, because of the irresponsibility in which it is exchanged, and its accepted lack of warrant, and because of the way gossip is dispersed and held within the community of gossip exchange, floats free from responsible counter-statement.

Gossip has no foundation or warrant that can be readily examined. It is irresponsible speech. A rumor may be true and you may believe it, but your belief is no justification of its truth. Gossip is not about true belief, but about social dynamics. It is constrained only by its means of transmission and the dispositions of individuals within a community to accept it.

We have argued that online information diffusion matches the conceptual patterns that define gossip. If online information diffusion is understood as a form of gossip, then we can understand that diffusion as a function of various social imperatives, shaping, reinforcing, undermining communities and individuals. The veracity of the information is often beside the point. Online social networks have become kingdoms of gossip, not of fact or truth.

We have described the essential properties of gossip both in general and online. These properties characterize something about the information shared and the social practices and behaviors relative to which the sharing of information gains its meaning and significance. Consequently, what we call the discursive grammar of gossip provides targets and topics for analysis. We offer three examples of topics.

[1] The diffusion of information represents a kind of investment in that information, producing a kind of temporary gossip-economy. If information diffusion is understood as a form of such a gossip-economy, then, of course, the veracity of the information is not critical, its value is.

[2] Information does not just spread through a network of friends, it reconstitutes a community though its sharing and resharing. The highly mediated and restricted nature of online interactions produces a particularly unstable and limited social configuration (community). Are such social confirmations replacing offline communities in some significant way?

[3] When information is exchanged through and as gossip, and since it is not self-replicating, the processes producing its diffusion maybe like the process of species diffusion into new environments. Online social networks and the communities of gossip-friends that they facilitate create a complex and dynamic communal, yet dispersed ecosystem. Information (including misinformation) spreads into and through that ecosystem in a way that is similar to the spread of a species into a new environment. Such a species diffusion is not like a virus. It is highly contingent, and depends on the dynamic interaction between the species (the information) and the ecosystem (individuals and communities).

All three of these conceptualizations of information diffusion highlight the fact that diffusion data reveals human data (separating out automated methods of diffusion). Information diffusion models should be understood as tools with which to explore the sociology of evolving online communities in conjunction with offline communities. We cannot help but wonder what kind of commitments, trust, common action, and exchange will be possible or encouraged by modern communities founded through and on online gossip and rumor. This question is critically important to consider and answer if we are to survive the negative effects of online social media culture.

## Disclosure Statement

The authors have no disclosures to share for this manuscript.